# Preserving the Basic Property of Stable Matching by Deleting a Pair

Ekta Gupta, Kalyani and Nitin*
Computer Science Engineering and Information Communications Technology
Jaypee University of Information and Technology
Waknaghat, Dist: Solan (H.P.), India

## ABSTRACT
This paper describes the transition of a male-pessimal matching set to optimal when it is a man-oriented approach by deleting a pair from matching set considering the score based approach. A descriptive explanation of the proposed algorithm both in a sequential and parallel manner is given. The comparison based theoretical analysis shows that the best case of the algorithm is lower bound of $n^3$.

## 1. INTRODUCTION
David Gale and Lloyd Shapley introduced the problem of stable matching in 1962 in a paper entitled "College Admissions and the Stability of Marriage. The basic algorithm states people to be heterosexual with n-men and n-women providing their preference lists for the opposite gender and hence the matching being 1:1. In some instances it came out to be a fact that man-oriented approach led to a man-pessimal matching where they did not get their best possible partner contrary to what Gale-Shapely proposed which leads to a worst case scenario. Though it has already been experimentally proved that the chances of having a worst case scenario for stable matching is extremely low, but occurrence of it prevents the parallel algorithm for stable matching to run, which could have reduced the running time dramatically.

This paper targets the above issue and solves it by deleting one pair from the already formed matching set using the score-based approach. In section 2 the related done so far in this field has been addressed. Section 3 deals with the basic concepts and notations that we have used in this paper. Section 4 depicts the proposed algorithm and section 5 shows the implementation details of the Modified GSA. Section 6 deals with the concluded results and finally section 7 discusses the future scope of the problem.

## 2. RELATED WORK
Much amount of work has been done in the field of Stable Matching since 1962, since the two economists David Gale and Lloyd Shapely first introduced it, [1] to today when they are awarded Nobel Prize in 2012 for the theory of stable allocations and the practice in market design and is still going on. A recent research [2] has emphasized on each and every aspect of Stable Matching Problem (SMP), such as incomplete lists, blocking pairs, ties etc. and has enlisted the research till now done to eradicate it. The time complexity of the basic GSA is given by $O(n^2)$, and to reduce it various researchers have tried their best. [3] has focused upon the number of ways an entity in one side of the partition can prefer the entity on the other side and has addressed various issues accordingly. It has considered the worst case specifically where the proposing party gets its worst possible partner going through (n-1) iterations for each of the 'n' members giving a time complexity of $O(n^2)$ as said by GSA and has further tried to reduce it by $O(2n\sqrt{n})$ using Latin Square matrices. As GSA is basically either man-oriented or woman oriented, depending upon the side taking its first step to propose, an approach [4] has been put forward to achieve optimality in the satisfaction of entities at both sides known as the egalitarian solution, has also been proposed which had a time complexity of $O(n^4)$. As we can see the time complexity achieved here is quite high, therefore [5] took an initiative to let the non-proposing party deceive GSA by changing its preference list to get the partner of choice hence achieving happiness. The disadvantage here was that the proposing party can be at stake of losing happiness. To give a solution to this [6] gave an approach where the men can deceive too. The authors of [7-10] has focused upon the basic algorithm and structure of GSA in a parallel way and the time complexity is found out to be $(n^2-2n+\log_2 n)$ which is less than $(n^2)$. But the parallel algorithm fails to execute in the worst case instances. To curb this limitation, [11] proposed a approach to change to change the worst case problem instance to a basic one by making minimal changes to the preference list of the entities. In our paper we present another approach to do the above by deleting the least happy pair using a score based approach. The details of the approach have been discussed in the further sections.

## 3. BASIC CONCEPTS AND NOTATIONS
Basically, the stable matching problem considers two sets M and W each of size n. M is a matrix of n-men along with the respective preference list for women. Similarly, W is a matrix of n-women along with the respective preference list for men. Here, we are considering the preference lists to be complete and strictly ordered [2] [3]. A complete list is such that a man needs to specify the ranks for all of his partners that are participating in the game, and a strict ordering of lists puts a bound that man needs to be clear about his thoughts for the preferences of his partners and therefore he cannot assign the same rank to more than one partner. In any instance of the matchmaking problem we uniquely match each man in set M with its woman (partner) in the set W for a man-oriented approach and vice versa if it is a woman-oriented approach, which means that GSA is partial. Either it favors men leading to a man optimal and woman pessimal solution or the other way round. To achieve global optimality with respect to both the sides we have egalitarian approach with a time complexity, $O(n^4)$.

Given a problem-instance, Ii a matching ŋ is a pairing of man Mi to woman Wj. If (M1, W3) belongs to ŋ, then we can say that M1 and W3 is a couple. A complete 1:1 matching of each man in set M to each woman in set W uniquely is known as a marriage. If a man and a woman in different couple in the matching set ŋ prefers to each other to their present partner then we say we have a blocking pair and the marriage is not stable [1] [3] [4]. Therefore, a stable matching is a marriage with no blocking pairs. Rank refers to the priority (position) in a person's preference list of his or her partner. We will denote the rank of woman Wj for man Mi in his preference list as $RM_i(W_j)$. In this paper we are considering score as the sum of all the ranks in ŋ, denoted by $Sŋ(I)$.

In the basic Gale Shapely algorithm it has been clearly specified that, in a man-oriented stable matching, it should always be man-optimal. But in some instances, we have men not assigned their highest possible ranked partners leading to a man-pessimal matching. For such instances merit of the Gale Shapley algorithm cannot be used efficiently [7] [8] [9]. Therefore, [11] has tried to eradicate the worst case scenario in $O(n^3)$ time, by making minimal changes in the preference list of the contestants. In paper we are trying to reduce this time complexity to some extent by achieving a lower bound of $(n^3)$ by considering an alternative score based approach.

Here we illustrate the scenario with the help of an example. Suppose we have set of 4 men and 4 women. Each preference list is ordered in increasing order from left to right as shown in TABLE 1. Lower the priority higher is the preference.

**Table 1. Set of 4 men and women along with their respective preference lists**

| $Man_i$ | Pref. List ($PL_{Mi}$) | $Woman_j$ | Pref. List ($PL_{Wj}$) |
|---|---|---|---|
| $M_1$: | $W_1, W_2, W_3, W_4$ | $W_1$: | $M_2, M_1, M_3, M_4$ |
| $M_2$: | $W_3, W_1, W_2, W_4$ | $W_2$: | $M_1, M_2, M_3, M_4$ |
| $M_3$: | $W_2, W_4, W_3, W_1$ | $W_3$: | $M_4, M_3, M_1, M_2$ |
| $M_4$: | $W_2, W_3, W_4, W_1$ | $W_4$: | $M_3, M_2, M_1, M_4$ |

If we follow the general GSA[1] we will end up having:

ŋ = {(M1, W2), (M2, W1), (M3, W4), (M4, W3)}

As we can see in the men-table (TABLE 2) we have all the men getting paired up with their second preferences, and the women get the men from their first preferences (TABLE 3).

**Table 2. Assignment of women for men in the men-table**

| $Man_i$ | Pref. List ($PL_{Mi}$) |
|---|---|
| $M_1$: | $W_1, W_2, W_3, W_4$ |
| $M_2$: | $W_3, W_1, W_2, W_4$ |
| $M_3$: | $W_2, W_4, W_3, W_1$ |
| $M_4$: | $W_2, W_3, W_4, W_1$ |

**Table 3. Assignment of women for men in the women-table**

| $Woman_j$ | Pref. List ($PL_{Wj}$) |
|---|---|
| $W_1$: | $M_2, M_1, M_3, M_4$ |
| $W_2$: | $M_1, M_2, M_3, M_4$ |
| $W_3$: | $M_4, M_3, M_1, M_2$ |
| $W_4$: | $M_3, M_2, M_1, M_4$ |

Gale Shapely Algorithm says that in case of a man-oriented approach a man always gets its best possible partner and a woman its worst possible partner i.e. it should be man-optimal and woman-pessimal. But the result we got in this case is a man-pessimal and woman-optimal solution. As the result shows, we have women happier than men when men initiate the proposal, contrary to what GSA says. This should not happen unless any woman cheats [5] by changing her preference list after anticipating the men's order of proposals and choosing the cheating strategy for herself to get the man she desires. But in this case the women gain a heavier side of the balance as the men (proposing entity) cannot do anything to save them from deception; therefore [6] proposed the cheating by men approach.

## 4. MODIFIED GALE SHAPELY ALGORITHM ($MOD_{GSA}$)

The above result we got is the worst case scenario where the proposing party who is expected to be happier than the non-proposing one is sad rather. It has been proved that if there are 16 men and 16 women then the probability that the worst case occurs is $10^{-45}$, which is very low [7]. Parallel algorithm is based upon divide and conquer principle having a time complexity of $n^2-2n+ [\log n]$ as stated in [8], which is better than the time complexity of basic Gale-Shapely Algorithm. But, parallel algorithms do not work for the worst case. We can avoid such worst case scenario to some extent by following the Modified GSA proposed in this paper.

In this algorithm we are taking as input a matrix of M for men and W for women, and their preference lists ordered according to their priority. We apply GSA on the basic problem instance $I_o$ and we denote the matching set found, by $ŋ_0$. For say, we have a set of 4 men and 4 women then the matching set formed will have 4 pairs with each man paired with his respective partner, we then denote the matching set $ŋ_0=[p_1, p_2, p_3, p_4]$ where $p_i$ denotes a pair i. Therefore, we can say that the matching set is a matrix with n-pairs for n being the size of the problem. Though the problem size is considered n × n, but for simplicity we will consider it n throughout the paper.

The for-loop in the Modified Gale-Shapely Algorithm runs for each pair which is given by the problem size only i.e. n. For each pair we delete the pair first. Then we apply GSA to the new matrix set of (n-1) men and (n-1) women, leading to a matching set $ŋ_i$. Finally we calculate the score of $ŋ_i$, as $S_{ŋi}(I_i)$. We continue to do so for the entire pairs $p_i$ in the original GSA matching set, and select to delete the pair with the minimum score, $S_{min}$. Therefore, the GSA matching set retained now has the minimum score. In the matching set ŋ, the score is found as the sum total of all the ranks of the partners in the preference list of the proposing party.

**ALGORITHM: MODIFIED GSA (MOD$_{GSA}$)**

Input: The problem instance with the men matrix M and the women matrix W along with their preference lists.

Output: A man-optimal matching set, $m_i$ for a man-oriented approach.

Precondition: The problem instance should produce the worst case scenario.

1. Calculate the score for the problem instance $I_0$
2. **For** each pair in matching set $m_0$ do
3.     Delete the pair $p_i$
4.     Form the matching set $m_i$
5.     Calculate the score, $S_m(I_i)$
6. **End for**
7. Delete the pair $p_i$ for which the score, $S_m$ is minimum
8. Output the matching set $m_i$ for which the pair has been deleted.

**Time Complexity:** The time complexity $T(n)_{GSA}$ of the above algorithm is given by $O(n^3)$.

**Proof of Complexity or Correctness:** The Gale Shapely algorithm takes $O(n^2)$ for a problem size n. We have the for-loop run for each pair. For a problem size n we always end up having n pairs. Therefore, we get the complexity to be calculated as:

Time Complexity, $T(n)_{GSA}=(n-1)^2 \times n$

$$=O(n^3)$$

We can improve this time complexity by following parallel GSA (MOD$_{P-GSA}$) instead of basic GSA at line 4. At line 1 we will follow the basic GSA. Here we have made possible for the parallel algorithm to execute successfully with high probability in case of a worst case scenario by deleting or ignoring one couple from the matching set $m_0$.

## 5. IMPLEMENTATION

In this section of our paper we will describe the implementation details of our algorithms both for MOD$_{GSA}$ and MOD$_{P-GSA}$. Then we have done a comparative analysis of these algorithms taking number of steps as the parameter. At a later stage we have represented our results graphically for both space complexity (number of steps) and time complexity (T(n)). Finally, we analyze the performance metric enhancement for MOD$_{P-GSA}$ w.r.t. MOD$_{GSA}$.

## 5.1 Detailed Explanation (MOD$_{GSA}$)

The Here we will consider the TABLE I and based upon it we will describe our algorithm for MOD$_{GSA}$ in a stepwise manner. We will consider the Modified Parallel GSA denoted by MOD$_{P-GSA}$ in the next section of implementation.

For problem instance in TABLE 1 the following steps describe the flow of the algorithm.

Step 1: Applying GSA we get the matching $m_0 = \{(M_1, W_2), (M_2, W_1), (M_3, W_4), (M_4, W_3)\}$ having the score, $S_m(I_o)= 2+2+2+2=8$.

The score we will calculate at a later stage should come less than this as we are trying to maximize happiness for men. This constraint would verify the correctness of our algorithm as less is the score more is the happiness.

Step 2: For deletion, we need to consider each pair in M. As for a problem size n we always end up having n-pairs, this for loop will run for n-times.

Step 3: We proceed first by considering $m_0[0]$ i.e. $(M_1, W_2)$. Now we are left with the matrix:

**Table 4. Reduced Table**

| Man$_i$ | Pref. List (PL$_{Mi}$) | Woman$_j$ | Pref. List (PL$_{Wj}$) |
|---|---|---|---|
| $M_2$: | $W_3, W_1, W_4$ | $W_1$: | $M_2, M_3, M_4$ |
| $M_3$: | $W_4, W_3, W_1$ | $W_3$: | $M_4, M_3, M_2$ |
| $M_4$: | $W_3, W_4, W_1$ | $W_4$: | $M_3, M_2, M_4$ |

Step 4: Applying GSA, $m_1 = \{(M_2, W_1), (M_3, W_4), (M_4, W_3)\}$

Step 5: For the above reduced problem instance $I_1$, score is given by, $S_m(I_1)= 5$

Step 6: Similarly doing it for all other pairs in $m$, we have

Delete $(M_2, W_1)$: $m_1=\{(M_1, W_2), (M_3, W_4), (M_4, W_3)\}$

$$S_m(I_2)= 6$$

Delete $(M_3, W_4)$: $m_1 =\{(M_1, W_1), (M_2, W_3), (M_4, W_2)\}$

$$S_m(I_3)= 3$$

Delete $(M_4, W_3)$: $m_1 = \{(M_1, W_1), (M_2, W_2), (M_3, W_4)\}$

$$S_m(I_4)= 4$$

Step 7: Choosing the pair with minimal score we delete $(M_3, W_4)$, and we are left with the matrix:

**Table 5. Result Table**

| Man$_i$ | Pref. List (PL$_{Mi}$) | Woman$_j$ | Pref. List (PL$_{Wj}$) |
|---|---|---|---|
| $M_1$: | W$_1$, $W_2, W_3$ | $W_1$: | $M_2$, M$_1$, $M_4$ |
| $M_2$: | W$_3$, $W_1, W_2$ | $W_2$: | $M_1, M_2$, M$_4$ |
| $M_4$: | W$_2$, $W_3, W_1$ | $W_3$: | $M_4, M_1$, M$_2$ |

The TABLE V clearly shows that now the men get their first preferences and women either their second or third, resulting into a man-optimal and therefore woman pessimal solution.

As we can see here, our problem size has been reduced to (n-1), but as we go on increasing the value of n, this hardly matters, if by doing so we get an overall happiness and preserve the basic property of Gale-Shapely Algorithm by reducing the occurrence of a worst case.

## 5.2 Modified Parallel GSA (MOD$_{P-GSA}$)

As cited in [9] [10], parallel GSA follows divide and conquer principle to solve the matchmaking problem in a parallel way taking $(n^2-2n+\log_2 n)$ steps, where n is the size of the main problem. As the name indicates, this has got two phases i.e. the division phase and the conquering (merging) phase. The division of the problem into sub-problems led to a tree like structure and problems at the same tree level are solved in a parallel fashion to produce a partial matching set which is then merged to form a higher level match. The conflict where a single man is matched twice at the same level with two women is solved by consulting the women's preference list. This whole process continues until we get the final result.

As cited in [9], parallel GSA does not work is a worst case scenario, therefore the input to $MOD_{P\text{-}GSA}$ i.e. the matching set $m_0$ is calculated following the $GSA_{BASIC}$, which takes at most $n^2$ number of steps in a worst case. Inside the algorithm where we obtain the matching set $m_i$ we will use the parallel GSA. Even here there is a chance that the worst case scenario may occur. But it has already been stated [8] that the chances of the occurrence of worst case are very rare and the rarity increases even more as we are searching for a worst case within the worst case.

**Time Complexity**: The time complexity, $T(n)_{P\text{-}GSA}$ of the Modified Parallel GSA is given by $O(n^3)$.

**Proof of Complexity or Correctness**: The Parallel Gale Shapely algorithm takes $(n^2-2n+\log_2 n)$ number of steps for a problem size n. As for-loop runs after we delete a pair, the problem size reduces to (n-1). This for-loop runs for each pair. For a problem size n we always end up having n pairs. Therefore, we get the complexity to be calculated as:

Time Complexity, $T(n)_{P\text{-}GSA} = ((n-1)^2 - 2(n-1) + \log_2(n-1)) \times n$

$= O(n^3)$

## 5.3 Comparison ($MOD_{GSA}$ 'vs' $MOD_{P\text{-}GSA}$)

The time complexity of both the algorithms $MOD_{GSA}$ and $MOD_{P\text{-}GSA}$ is given by $O(n^3)$ when calculated. Taking the worst case scenario as input and the number of steps required as the parameter here we have done a theoretical comparative analysis. Finally we have deduced the performance enhancement for MODP-GSA in comparison to MODGSA and we concluded that as the value of n increases the performance enhancement metric goes on giving better results.

**Table 6. Comparative Analysis**

| NUMBER OF PEOPLE IN EACH SET (n) | NUMBER OF STEPS | | | PERFORMANCE ENHANCEMENT IN $MOD_{P\text{-}GSA}$ W.R.T. $MOD_{GSA}$ |
|---|---|---|---|---|
| | GSA | $MOD_{GSA}$ | $MOD_{P\text{-}GSA}$ | |
| 3 | 9 | 12 | 6 | LOW |
| 4 | 16 | 36 | 20 | LOW |
| 5 | 25 | 80 | 55 | LOW |
| 6 | 36 | 150 | 108 | INTERMEDIATE |
| 7 | 49 | 252 | 189 | INTERMEDIATE |
| 8 | 64 | 392 | 304 | INTERMEDIAE |
| 9 | 81 | 576 | 468 | INTERMEDIATE |
| 10 | 100 | 810 | 670 | INTERMEDIATE |
| 11 | 121 | 1100 | 924 | HIGH |
| 12 | 144 | 1452 | 1236 | HIGH |
| 13 | 169 | 1872 | 1612 | HIGH |
| 14 | 196 | 2366 | 2058 | HIGH |
| 15 | 225 | 2940 | 3184 | HIGH |
| 16 | 256 | 3600 | 3184 | HIGH |

## 5.4 Graphical representation

Representing the data from TABLE 6 in a graphical form we obtain FIGURE 1 showing the comparison of performance for each algorithm. Here also we can see that the difference between the peak points for $MOD_{GSA}$ and $MOD_{P\text{-}GSA}$ keeps on increasing as the value of n increases.

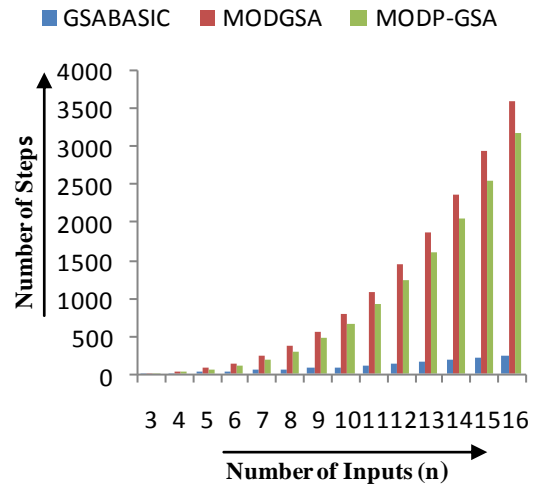

**Figure 1: Comparison graph considering 'No. of Steps' as a parameter**

We know that the number of steps required for an algorithm to run is directly proportional the time it will take to run on any machine. When we run the algorithm for various problem instances we obtained the graph given in FIGURE 2. The graph shows the variation in time complexities for $MOD_{GSA}$ and $MOD_{P\text{-}GSA}$ and the edge $MOD_{P\text{-}GSA}$ obtains over $MOD_{GSA}$ for larger values of n.

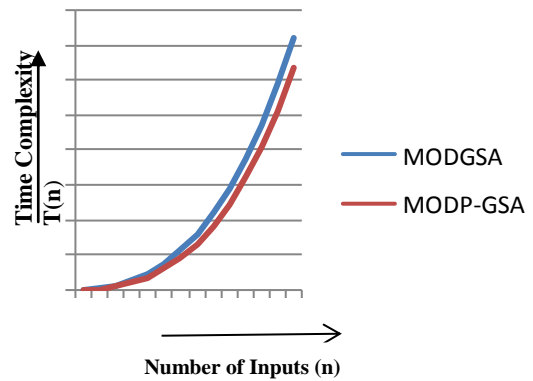

**Figure 2: Comparison of Time Complexities of $MOD_{GSA}$ AND $MOD_{P\text{-}GSA}$**

## 6. CONCLUSION

This paper explores the worst case scenario of Gale Shapely Algorithm (GSA) and improves it by deleting one pair to achieve greater happiness. For this we have proposed an algorithm, $MOD_{GSA}$, based on $GSA_{BASIC}$ which takes $O(n^3)$ time. Further, we have been trying to decrease this time complexity, by following the parallel GSA ($GSA_{PARALLEL}$), denoted by $MOD_{P\text{-}GSA}$. We have taken a number of steps to run the algorithm as our parameter and represented our results both in tabular and graphical form. However, on comparing we deduced the result that $MOD_{P\text{-}GSA}$ gives better performance than $MOD_{GSA}$ for higher values of n and hence achieving greater stability.

## 7. FUTURE WORK

In future we are looking forward to reduce the time complexity achieved i.e. $O(n^3)$ by running the algorithm on different processors and taking processor specific characteristics into consideration. $MOD_{P-GSA}$ fails if we encounter a worst-case within our basic problem instance $I_0$, and we would try to eradicate such a situation. In our future work we will also put some light upon the completeness of the algorithm and how ties and incomplete lists affect this. We will also consider the cheating of women and its effect on the $MOD_{GSA}$ as well as on $MOD_{P-GSA}$.

## 8. ACKNOWLEDGMENT

We are grateful to Nitin for his valuable comments and his efficient use of time for painstakingly checking the correctness of proof of complexity in the preliminary versions of this paper. We would also like to acknowledge Jaypee University of Information Technology for its continual support both financially and technically to help improve the presentation of this paper. Finally, we would like to thank the anonymous referees, for their constructive comments that greatly helped in the improvement of this paper.